\documentclass[12pt]{article}

\catcode`@=12
\def\nd{\nu \phantom{} DE}
\def\aa{{\cal A}}
\begin{document}
\thispagestyle{empty}
\begin{flushright}
UCRHEP-T405\\
February 2006\
\end{flushright}
\vspace{0.5in}
\begin{center}
{\LARGE \bf Connecting Dark Energy to Neutrinos\\
with an Observable Higgs Triplet\\}
\vspace{1.0in}
{\bf Ernest Ma\\}
\vspace{0.2in}
{\sl Physics Department, University of California, Riverside,
California 92521, USA\\}
\vspace{0.5in}
{\bf Utpal Sarkar\\}
\vspace{0.2in}
{\sl Physical Research Laboratory, Ahmedabad 380009, India\\}
\vspace{1.0in}
\end{center}

\begin{abstract}\
To connect the scalar field (acceleron) responsible for dark energy to
neutrinos, the usual strategy is to add unnaturally light neutral singlet
fermions (right-handed neutrinos) to the Standard Model.  A better choice
is actually a Higgs triplet, through the coupling of the acceleron to the
trilinear Higgs triplet-double-doublet interaction.  This hypothesis
predicts an easily observable doubly-charged Higgs boson at the forthcoming
Large Hadron Collider (LHC).
\end{abstract}

\newpage
\baselineskip 24pt
The existence of dark energy \cite{dark} may be attributed to a scalar field
called the acceleron (or quintessence) \cite{quint} whose equation of motion
involves a term of negative pressure, allowing the present Universe to expand
at an accelerated rate.  The acceleron may also form a condensate and couple
to matter in such a way that the observed neutrino masses are dynamical
quantities.  This is the scenario of mass varying neutrinos \cite{vary},
motivated by the proximity of the effective mass scale of dark energy to
that of neutrinos, which may have some interesting consequences \cite{kap,lg}.

To make the connection, the usual strategy is to introduce 3 right-handed
neutrinos $N_i$, i.e. 3 neutral fermion singlets under the electroweak
$SU(2)_L \times U(1)_Y$ gauge group.  However, contrary to the cherished
expectation that $m_{N_i}$ should be very large (thereby triggering the
canonical seesaw mechanism \cite{seesaw} and yielding naturally small
Majorana neutrino masses $m_{\nu_i}$), they have to be very small, i.e.
of order eV, to be compatible with dark energy.  In view of this problem,
alternative mechanisms for the origin of $m_{\nu_i}$ should be explored
\cite{models}.

In the Standard Model, naturally small Majorana neutrino masses come from
the unique dimension-five operator \cite{w79}
\begin{equation}
{\cal L}_{eff} = {f_{ij} \over \Lambda} (\nu_i \phi^0 - l_i \phi^+)
(\nu_j \phi^0 - l_j \phi^+) + H.c.,
\end{equation}
which can be realized at tree level in exactly 3 ways \cite{m98}, one of
which is of course the canonical seesaw mechanism with 3 right-handed
neutrinos. Another way is to add a Higgs triplet \cite{triplet}
\begin{equation}
\Delta = \pmatrix{\xi^+/\sqrt 2 & \xi^{++} \cr -\xi^0 & -\xi^+/\sqrt 2}
\end{equation}
with trilinear couplings to both the lepton doublets $(\nu_i,l_i)$ and the
Higgs doublet $\Phi = (\phi^+,\phi^0)$, i.e.
\begin{equation}
{\cal L}_{int} = f_{ij} [\nu_i \nu_j \xi^0 + {1 \over \sqrt 2} (\nu_i l_j +
l_i \nu_j) \xi^+ + l_i l_j \xi^{++}] + \mu \Phi^\dagger \Delta \tilde \Phi
+ H.c.,
\end{equation}
where $\tilde \Phi = (\bar \phi^0,-\phi^-)$.  As a result \cite{ms98},
\begin{equation}
({\cal M}_\nu)_{ij} = {2 f_{ij} \mu \langle \phi^0 \rangle^2 \over
m^2_{\xi^0}}.
\end{equation}
If $\mu = \mu(\aa)$, i.e. a function of the acceleron field $\aa$, then this is
in fact a natural realization of mass varying neutrinos with $m_\xi$ of order
the electroweak scale.

In all previous proposals of neutrino mass with a Higgs triplet, there is
no compelling reason for $m_\xi$ to be this low.  One possible exception
\cite{mrs00} is the case of large extra space dimensions, where $m_\xi$
should be below whatever the cutoff energy scale is, but that is only a
phenomenological lower bound.  On the other hand, if  dark energy is indeed
connected to neutrinos through the Higgs triplet, then at least $\xi^{++}$
will be unambiguously observable at the forthcoming Large Hadron Collider
(LHC).

Consider the most general Higgs potential consisting of $\Phi$ and $\Delta$,
i.e.
\begin{eqnarray}
V &=& m^2 (\Phi^\dagger \Phi) + M^2 (Tr \Delta^\dagger \Delta) + {1 \over 2}
\lambda_1 (\Phi^\dagger \Phi)^2 + {1 \over 2} \lambda_2 (Tr \Delta^\dagger
\Delta)^2 \nonumber \\
&+& {1 \over 2} \lambda_3 (Tr \Delta^\dagger \Delta \Delta^\dagger \Delta)
+ \lambda_4 (\Phi^\dagger \Phi)(Tr \Delta^\dagger \Delta) + \lambda_5
(\Phi^\dagger \Delta^\dagger \Delta \Phi) \nonumber \\
&+& \mu(\tilde \Phi^\dagger \Delta^\dagger \Phi + \Phi^\dagger \Delta \tilde
\Phi). \label{scalar}
\end{eqnarray}
Let $\langle \phi^0 \rangle = v$ and $\langle \xi^0 \rangle = u$, then
\begin{eqnarray}
&& v[m^2 + \lambda_1 v^2 + \lambda_4 u^2 - 2 \mu u] = 0, \\
&& u[M^2 + (\lambda_2+\lambda_3) u^2 + \lambda_4 v^2] - \mu v^2 = 0.
\end{eqnarray}
For $|\mu| << |m|, |M|$, and
\begin{equation}
m^2<0, ~~~ \lambda_1 M^2 - \lambda_4 m^2 > 0,
\end{equation}
we have the unique solution
\begin{equation}
v^2 \simeq -{m^2 \over \lambda_1}, ~~~ u \simeq {\mu v^2 \over M^2 +
\lambda_4 v^2}.
\end{equation}
The Higgs triplet masses are then
\begin{eqnarray}
m^2_{\xi^{++}} &\simeq& M^2 + (\lambda_4 + \lambda_5) v^2, \\
m^2_{\xi^+} &\simeq& M^2 + (\lambda_4 + \lambda_5/2) v^2, \\
m^2_{\xi^0} &\simeq& M^2 + \lambda_4 v^2.
\end{eqnarray}
Once produced, the decay of $\xi^{++}$ into two charged leptons is an
unmistakable signature with negligible background.  Its decay branching
fractions also map out $|f_{ij}|$, i.e. the entire neutrino mass matrix
up to an overall scale \cite{mrs00}.

In a model of neutrino dark energy ($\nd$), the neutrino mass $m_\nu$
is a dynamical quantity.  It is assumed to be a function of a
scalar field $\aa$ (the acceleron) with a canonically normalized kinetic
term and $\partial m_\nu / \partial \aa \neq 0$.  In the nonrelativistic
limit, $m_\nu$ depends on the total density $n_\nu$ of the thermal
background of neutrinos and antineutrinos, and the energy or effective 
potential of the system is given by
\begin{equation}
V = m_\nu n_\nu + V_0 (m_\nu).
\end{equation}
The thermal background and the scalar potential $V_0 (m_\nu)$
will act in opposite directions and at any instant of time, the
minimum of the effective potential is given by
\begin{equation}
V^\prime (m_\nu) = n_\nu + V_0^\prime (m_\nu) = 0.
\end{equation}
We assume the curvature scale of $V$ to be much larger than the Hubble
expansion rate, so that the adiabatic approximation is valid. In other
words, the solution of Eq.~(14) for $m_\nu$ is assumed to be valid 
instantaneously.

For an adiabatic expansion of the Universe, the density of matter varies
with the scale factor as
\begin{equation}
\rho \propto R^{-3(1+ \omega)} ,
\end{equation}
where $\omega$ is a time-independent parameter, which enters in the
following simple equation of state:
\begin{equation}
\rho (t) = \omega p(t) .
\end{equation}
In a $\nd$ model, it was shown that $\omega$ satisfies the equation
\begin{equation}
\omega + 1 = - {V^\prime (m_\nu) \over m_\nu V} =
{\Omega_\nu \over \Omega_\nu + \Omega_{DE}},
\end{equation}
where $\Omega_{DE} = \rho_{DE}/\rho_c$ is the contribution of $V_0(m_\nu)$
to the energy density and $\Omega_\nu = n_\nu/\rho_c$ is the
neutrino energy density.  Since the observed value \cite{dark}
$$ \omega = - 0.98 \pm 0.12 $$
is close to $-1$ at the present time, $\Omega_\nu$ should be much
less compared to $\Omega_{DE}$. These considerations restrict
the possibilities of the form of the potential.  For small $d \omega /
d n_\nu$, the variable mass of the neutrino is proportional to the
neutrino density to the power $\omega$:
$$ m_\nu \propto n_\nu^\omega .$$
The above general considerations are valid, independent of the details of the
particular model of neutrino mass.  However, most phenomenological
implications are specific to such details, with a few general
features which are common to all models \cite{kap}.

In the present scenario, for the effective neutrino mass to vary, we have
to associate the acceleron field $\aa$ with the trilinear coupling of $\Delta$
with $\Phi$, so that the effective neutrino mass becomes dependent on the
field $\aa$.  This simply means that we set $\mu = \mu(\aa)$ in
the scalar potential of Eq.~(5).  As for the self-interactions of $\aa$,
we may assume for example the following potential:
\begin{eqnarray}
V_0 &=&  \Lambda^4~ \log (1 + |\bar \mu / \mu(\aa)|).
\end{eqnarray}
Using Eq.~(4), the effective low-energy Lagrangian is then given by
\begin{equation}
- {\cal L}_{eff} = f_{ij} ~|\mu(\aa)| ~{\langle \phi^0 \rangle^2 \over
m^2_{\xi^0}}~ \nu_i \nu_j +H.c. + \Lambda^4 \log (1+ |\bar \mu /\mu(\aa)|),
\end{equation}
and Eq.~(13) is of the form
\begin{equation}
V(x) = ax + b \log \left(1+{c \over x}\right),
\end{equation}
where $x=m_\nu \propto |\mu(\aa)|$ and $a,b,c$ are all positive.  
For $4b/ac<<1$, $x_{min} \simeq b/a$, so
\begin{equation}
m_\nu \propto n_\nu^{-1},
\end{equation}
as desired. As a condition of naturalness, it has been argued that the 
mass of the scalar field should not be larger than the order of 1 eV and 
$\Lambda \sim 10^{-3}$ eV.  In the canonical realization of mass varying
neutrinos using right-handed neutrinos $N$, this would imply small 
$NN$ Majorana masses as well as tiny $\nu N$ Dirac masses, which are 
clearly rather unnatural.  Here, the requirement is simply that
$m_{\xi^0}$ be of order $\langle \phi^0 \rangle$, which is a much more
reasonable condition.

Thus the mass of $\xi^0$ is predicted to
be in the range of $80 - 500$ GeV. The lower limit is the
present experimental bound from the direct search of the triplet Higgs scalar,
while the upper limit comes from the requirement that it should not be
too large compared to the electroweak breaking scale,
otherwise it would be difficult to explain neutrino masses much
below 1 eV. The form of $\mu(\aa)$ was
discussed in the original paper \cite{vary} to be
$\mu(\aa) \sim \lambda \aa$
or $\mu(\aa) \sim \mu e^{\aa^2/f^2}$. We shall not go into the
details of this discussion on the dynamics of this model, although some
of the generic problems of mass varying neutrinos are common
to the present model as well \cite{peccei}.

Depending on the form of $\mu(\aa)$, global lepton number may be broken 
spontaneously in such a model of $\nd$, thereby creating a massless 
Goldstone boson, i.e. the Majoron.  However, as shown below, its 
coupling to ordinary matter is highly suppressed, hence its existence 
is acceptable phenomenologically.  If we take the case $\mu(\aa) \sim 
\lambda \aa$ (where $\aa$ is complex), we can express the field $\aa$ as
$$ \aa = {1 \over \sqrt{2} } (\rho + \sqrt 2 z) e^{i \varphi} $$
where $z$ is the vacuum expectation value or condensate of $\aa$.  Similarly,
\begin{equation}
\phi^0 = {1 \over \sqrt 2} (H + \sqrt 2 v) e^{i\theta}, ~~~~
\xi^0 = {1 \over \sqrt 2} (\zeta + \sqrt 2 u) e^{i\eta},
\end{equation}
with $v$ and $u$ as the vacuum expectation values of
$\phi^0$ and $\xi^0$ respectively. The longitudinal component of
the $Z$ boson ($G^0$), the physical Majoron ($J^0$)
and the massive combination ($\Omega^0$) of
$(z \varphi, u \eta, v \theta)$ are given by:
\begin{eqnarray}
G^0&=&\frac{v^2 \theta + 2 u^2 \eta}{\sqrt{v^2 + 4 u^2}}\,, \nonumber \\[.1in]
J^0 &=& {(v^2 + 4 u^2) z^2 \varphi + v^2 u^2 \eta - 2 u^2 v^2 \theta \over
\sqrt {z^2(v^2 + 4 u^2)^2 + u^2 v^4 + 4 v^2 u^4}} \nonumber \\[.1in]
\Omega^0&=&\frac{\varphi - \eta + 2 \theta}{\sqrt {z^{-2} + u^{-2} + 4 v^{-2}}}
\,,
\end{eqnarray}
respectively. The heavy $\Omega^0$ is almost degenerate
in mass with $\zeta$. They are essentially the reincarnations of $\xi^0$. 
The massless $J^0$ is potentially a problem phenomenologically but its 
couplings to all leptons are strongly suppressed by
$(u/v)^2$, and can safely be neglected in all present experiments.

Since the triplet Higgs scalars cannot be much heavier than 
the usual Higgs doublet, they should be observable at the LHC as well as the 
proposed future International Linear Collider (ILC). 
The phenomenology of such triplet Higgs scalars has been discussed in
\cite{mrs00}. The same-sign dileptons will be the most dominating
decay modes of the $\xi^{++}$. Complementary measurements of $|f_{ij}|$ at
the ILC by the process $e^- e^- (\mu^- \mu^-) \to l^-_i l^-_j$ would
allow us to study the structure of the neutrino mass matrix in
detail. Of course, these features are generic to any model with a Higgs 
triplet as the origin of Majorana neutrino masses.  The difference here is 
that it is also accompanied by the unusual predictions of mass varying 
neutrinos in neutrino oscillations \cite{kap,bhm05}.

In conclusion, we have pointed out in this paper that if the neutrino mass 
$m_\nu$ is dynamical and related to dark energy through the acceleron $\aa$, 
then the most natural mechanism for generating $m_\nu$ is that of the Higgs 
triplet, rather than the canonically assumed right-handed neutrino. 
The mass scale of the triplet Higgs scalars is predicted to be close to 
that of electroweak symmetry breaking, hence it has an excellent chance 
of being observed at the LHC and ILC.  Aspects of this model relating to 
cosmology and neutrino oscillations are similar to other existing models 
of dark energy.

This work was supported in part by the U.~S.~Department of Energy under
Grant No. DE-FG03-94ER40837.  EM thanks the Physical Reserach Laboratory, 
Ahmedabad, India for hospitality during a recent visit. We thank Bipin 
Desai for an important comment.

\bibliographystyle{unsrt}

\end{document}